\documentclass[12pt]{article}
\usepackage{epsfig}
\usepackage{amsmath,amssymb}

\def\beq{\begin{equation}}
\def\eeq#1{\label{#1}\end{equation}}
\def\eeqn{\end{equation}}

\def\beqa{\begin{eqnarray}}
\def\eeqa#1{\label{#1}\end{eqnarray}}
\def\eeqan{\end{eqnarray}}

\let\bar=\overbar

\def\Dslash{\not{\hbox{\kern-4pt $D$}}}
\def\dslash{\not{\hbox{\kern-2pt $\del$}}}

\def\msb{{\bar{\ssstyle M \kern -1pt S}}}

\def\Title#1{\begin{center} {\Large {\bf #1} } \end{center}}

\begin{document}

\Title{A proposal to solve some puzzles\\ in semileptonic $B$ decays}

\bigskip\bigskip

Proceedings of CKM 2012, the 7th International Workshop on the CKM Unitarity Triangle, University of Cincinnati, USA, 28 September - 2 October 2012\\ 
\bigskip

\begin{raggedright}  

{\it Sascha Turczyk\index{Turczyk, S.}\\
Ernest Orlando Lawrence Berkeley National Laboratory, University of California,\\ Berkeley, CA 94720, USA\\}
\bigskip
{\it Florian U.\ Bernlochner\\
University of Victoria,\\ Victoria, British Columbia, Canada V8W 3P\\}
\bigskip
{\it Zoltan Ligeti\\
Ernest Orlando Lawrence Berkeley National Laboratory, University of California,\\ Berkeley, CA 94720, USA\\}
%

\end{raggedright}

\section{Introduction}

About 25\% of all $B$ mesons decay semileptonically via the tree-level $b\to c$ quark transition. The study of such decays allow for the precision determination of the Cabibbo-Kobayashi Maskawa matrix element $|V_{cb}|$ and provide important background estimates or input for the study of rare decays. For the past ten years several puzzling features of this decay mode were observed, which individually are not several sigma problems, but remained unresolved.

In the quark model picture, if one considers $1S$, $1P$, and $2S$ excitations, in total eight mesons should exist: We refer to the two ground state charm mesons as $D^{(*)}$, the four orbitally excited as $D^{**}$, and the radial states as $D^{\prime(*)}$. Several tensions arise in decay modes involving the two broad members of the $1P$ excitations, either between different measurements or between measurements and theory predictions.  The relevant points for our discussion are:

\begin{figure}[htpb]
   \centering \includegraphics[width=0.35\columnwidth]{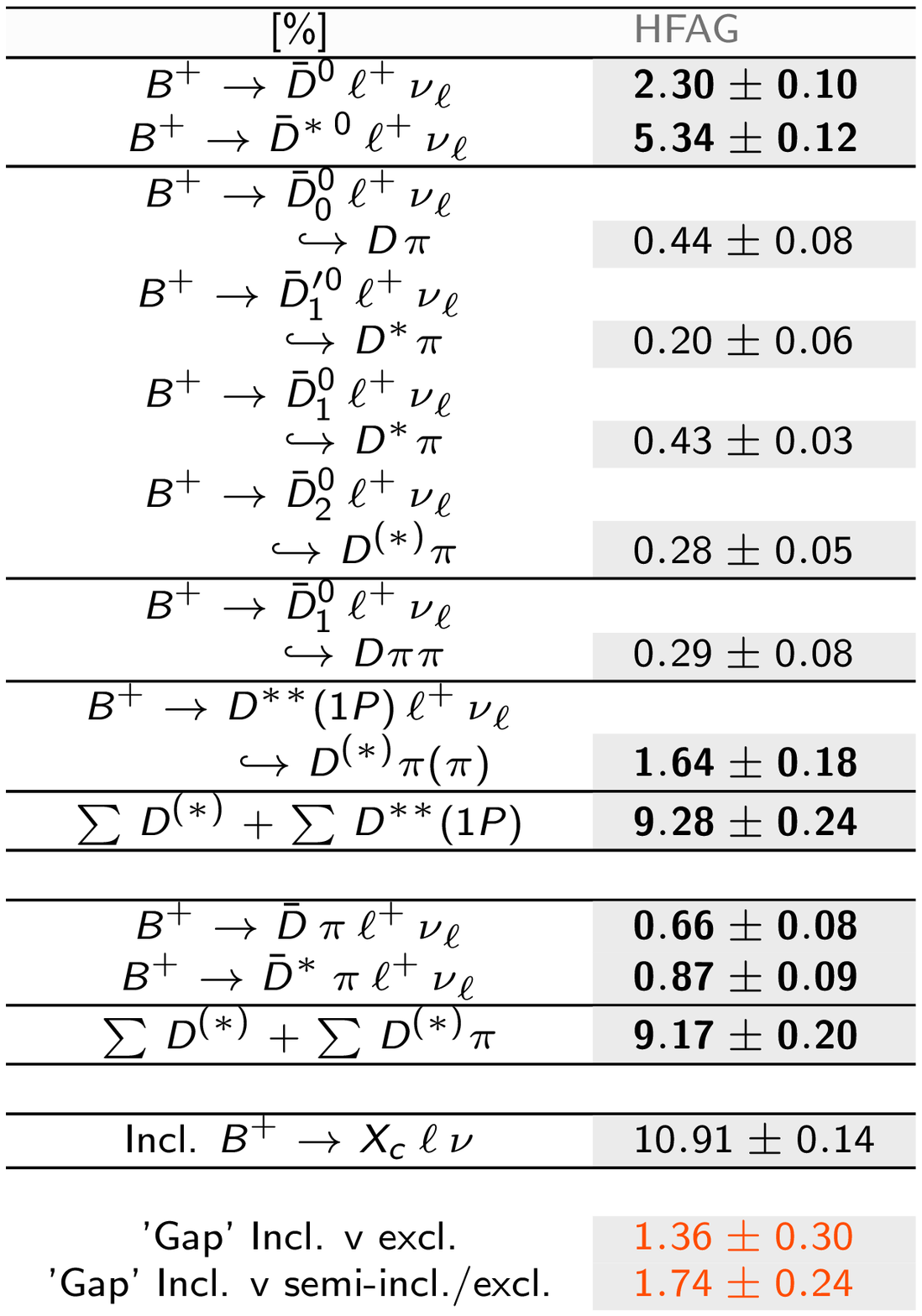}\hfill\includegraphics[width=0.5\columnwidth]{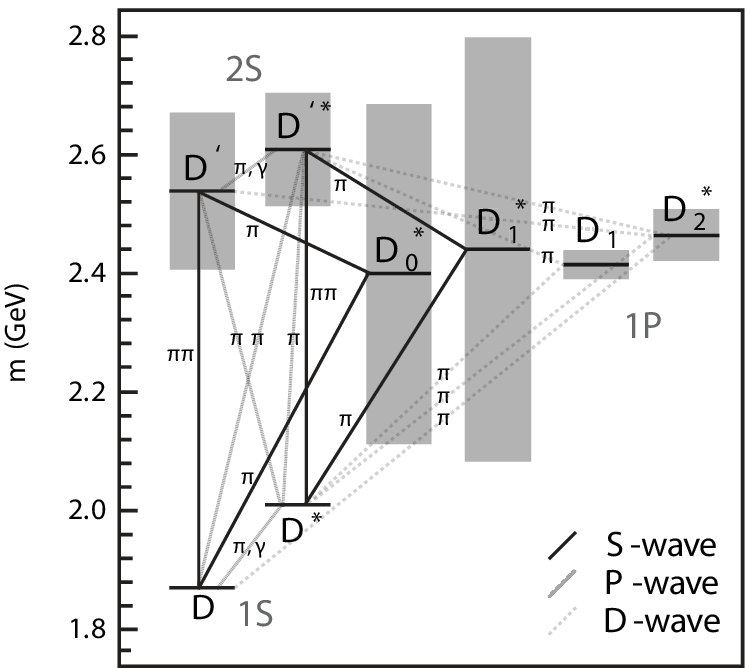}
    \caption{[Left] Branching fractions from HFAG~\cite{Asner:2010qj} and our private averages. For $B \to D^{(*)} \, \pi \, \ell \, \bar \nu_\ell$: weighted average, assuming a 100\% correlation, of both $B^{0,+}$ isospin modes. [Right] Strong decays of the $D^{\prime (*)}$ involving pion emissions, omitting possible near off-shell transitions with a $\rho$ and $\eta$. Grey bands correspond to the measured widths of the excited states. }\label{decay_modes}
\end{figure}

\begin{enumerate}\vspace*{-4pt}\itemsep 0pt
    \item The sum of the measured exclusive rates is less than the inclusive one, where the measurements are listed in Fig.~\ref{decay_modes}. Combining the measurements we find an absolute gap between the measurements, which should be compared to the branching fraction of about ${\cal O}(10\%)$. We have
    \begin{itemize}\vspace*{-4pt}\itemsep 0pt
	\item With the semi-inclusive $[\sum D^{(*)}+\sum D^{*}\pi] $ branching fractions the gap is \[(1.74 \pm 0.24)\%\]
	 \item The measured $1P$ decay $[\sum D^{(*)}+\sum D^{**} \to D^{(*)}\pi]$ amounts to a gap of \[(1.36 \pm 0.30 )\%\]
	 \item The quoted numbers slightly differ from the ones in our publication \cite{Bernlochner:2012bc}: a fraction of $B \to D_1 \, \ell \, \bar \nu_\ell$ with $D_1 \to D \pi \pi$ estimated by the ratio of the non-leptonic $B \to D_1 \pi $ with $D_1 \to D \pi \pi$ and $D_1 \to D^* \pi$ was added and the Belle lower limit on $B \to D_1' \, \ell \, \bar \nu_\ell$ was fully included. 
    \end{itemize}\vspace*{-4pt}
    \item The exclusive vs. inclusive determination of $|V_{cb}|$ differs as \cite{Beringer:1900zz}
	\begin{alignat*}{2}
		    |V_{cb}| &= (41.9 \pm 0.7) \times 10^{-3} &\quad &\text{(inclusive)} \\
		    |V_{cb}| &= (39.6 \pm 0.9) \times 10^{-3} &\quad &\text{(exclusive)} 
	        \end{alignat*}
    \item ``1/2 vs 3/2 puzzle'' \cite{Bigi:2007qp}: Theory prediction in conflict with data, see Fig.~\ref{decay_modes}
	    \begin{alignat*}{2}
			    \mathcal{B}(B^+ \to D^{**}_{1/2= \text{broad} } \, \ell^+ \, \nu ) /  \mathcal{B}(B^+ \to D^{**}_{3/2= \text{narrow}} \, \ell^+ \, \nu ) &\sim 0.1-0.2  &\quad &\text{[Theory]}\\
			    \mathcal{B}(B^+ \to D^{**}_{1/2= \text{broad} } \, \ell^+ \, \nu ) /  \mathcal{B}(B^+ \to D^{**}_{3/2= \text{narrow}} \, \ell^+ \, \nu ) &\sim 1 &\quad &\text{[Data]}\,.
		        \end{alignat*}
\end{enumerate}\vspace*{-4pt}

Although the size of the gap depends on the actual interpretation of the measurement and the used data, it remains significant in a statistical sense. In experimental analyses this gap is often filled up with a mix of known states, or the analyses are restricted to regions of phase space where decays making up this 'Gap' no longer contribute significantly due to kinematic restrictions. But this is unsatisfactory for many reasons and a thorough understanding of the matter is highly desirable. Here we  investigate the viability of a proposal, which could solve or at least ease some of these tensions.

\section{Proposal and its Viability}

The allowed decay modes of the considered excitations are displayed in Fig.~\ref{decay_modes}. The most important feature is, that the radially excited $2S$ mode can decay via a s-wave to the orbitally excited broad $1P$ modes, but only via d-waves into the narrow $1P$ states. The pion emitted in this strong decay has a soft momentum of $p_\pi \sim 0.01 - 0.5 \text{ GeV}$, allowing for the possibility of missing detection.  We investigate the possibility of a substantial branching fraction into radial states of the order
\begin{equation}
	    {\cal B}\big(B\to D^{\prime(*)}\ell\bar\nu\big) \sim {\cal O}(1\,\%)\,.
\end{equation}
If true, this could ease many of the above mentioned tensions in the following way:
\begin{enumerate}\vspace*{-4pt}\itemsep 0pt
	    \item It would be sufficient to saturate the inclusive rate closing the gap.
	    \item Decays involving the production of experimentally challenging soft pions could enhance the observed decay rate to the broad states $s_l^{\pi_l} = \frac12 ^+$ states, enhancing their population and thus ease the ''1/2 vs 3/2 puzzle``
	    \item The mass gap of the $1S$ and $2S$ is relatively small and thus the
		  charged lepton energy spectrum stays hard, which is in agreement with observations.
	    \item There is no direct conflict between the hypothesis and the ${\cal B}(B\to D^{(*)}\pi\ell\bar\nu)$ measurement: The $D^{\prime(*)}$ decay would yield two or more pions most of the time.
\end{enumerate}\vspace*{-4pt}

In order to investigate the viability, we want to estimate the possible branching fractions. The decay distributions are the same as for the ground-states $D^{(*)}$ up to different form~factors:
\begin{align}
	    \frac{\text{d}\Gamma_{D^{\prime*}}}{\text{d}w} &= 
  \frac{G_F^2 |V_{cb}|^2\, m_B^5}{48\pi^3}\, r^3 (1-r)^2\, \sqrt{w^2-1}\,
  (w+1)^2   \bigg[ 1 + \frac{4w}{w+1}\, \frac{1-2rw+r^2}{(1-r)^2} \bigg] 
  \big[F(w)\big]^2 \nonumber\\
 \frac{\text{d}\Gamma_{D^\prime}}{\text{d}w} &=
   \frac{G_F^2 |V_{cb}|^2\, m_B^5}{48\pi^3}\, r^3 (1+r)^2\, (w^2-1)^{3/2}\,
   \big[G(w)\big]^2\,.\label{rates}
\end{align}

The momentum transfer is small $1\leq w=v\cdot v^\prime \lesssim 1.3$ and thus we investigate a linear and a quadratic interpolation of the Isgur Wise function in the heavy quark limit. At zero recoil $\xi_2(1) = 0$ and the rate at $w=1$ comes entirely from $\Lambda_\text{QCD}/m_b$ corrections. We expect the slope to be positive, because the only change from $1S$ to $2S$ is, that the expectation value of the distance from the  heavy quark of a spherically symmetric wave function is increased. 
For the estimate we use the quark model estimate \cite{Ebert:1999ed}, hoped to be valid at $w = 1$. We further modify an existing light-cone sum rules calculation \cite{Faller:2008tr} to project out the ground-state, with the hope to be reasonably valid at $w_\text{max}$. 
\begin{figure}[htbp]
    \centering\includegraphics[width=0.45\columnwidth]{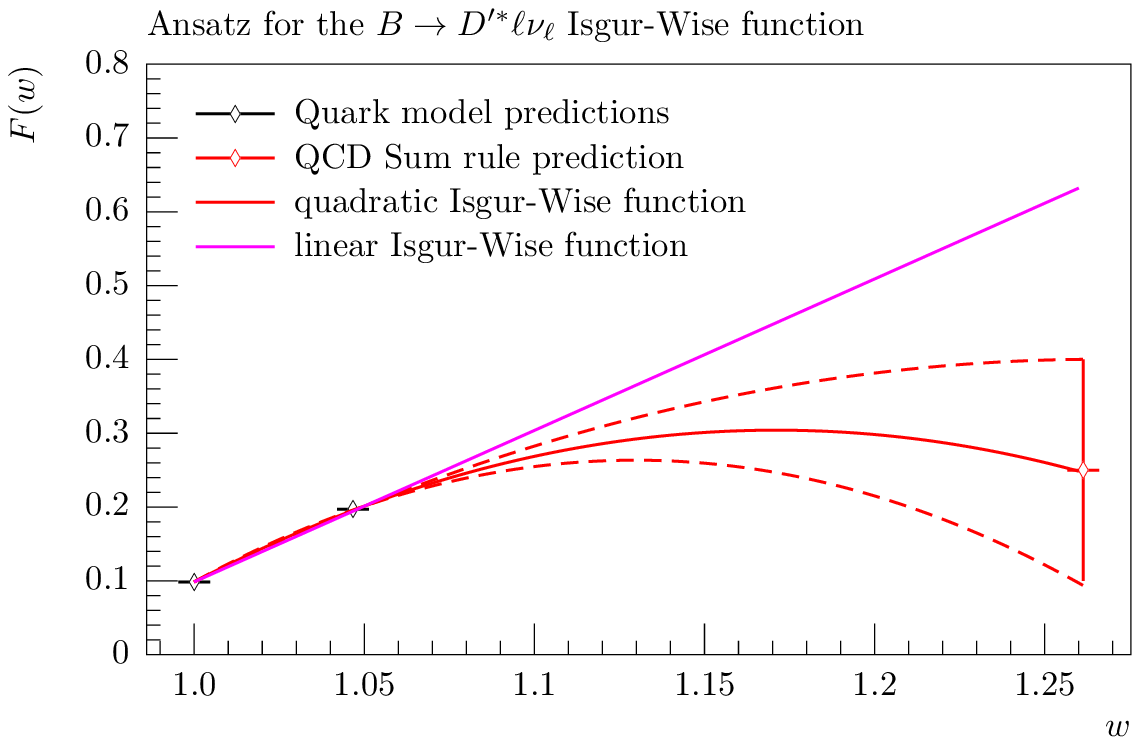}\hfill\includegraphics[width=0.45\columnwidth]{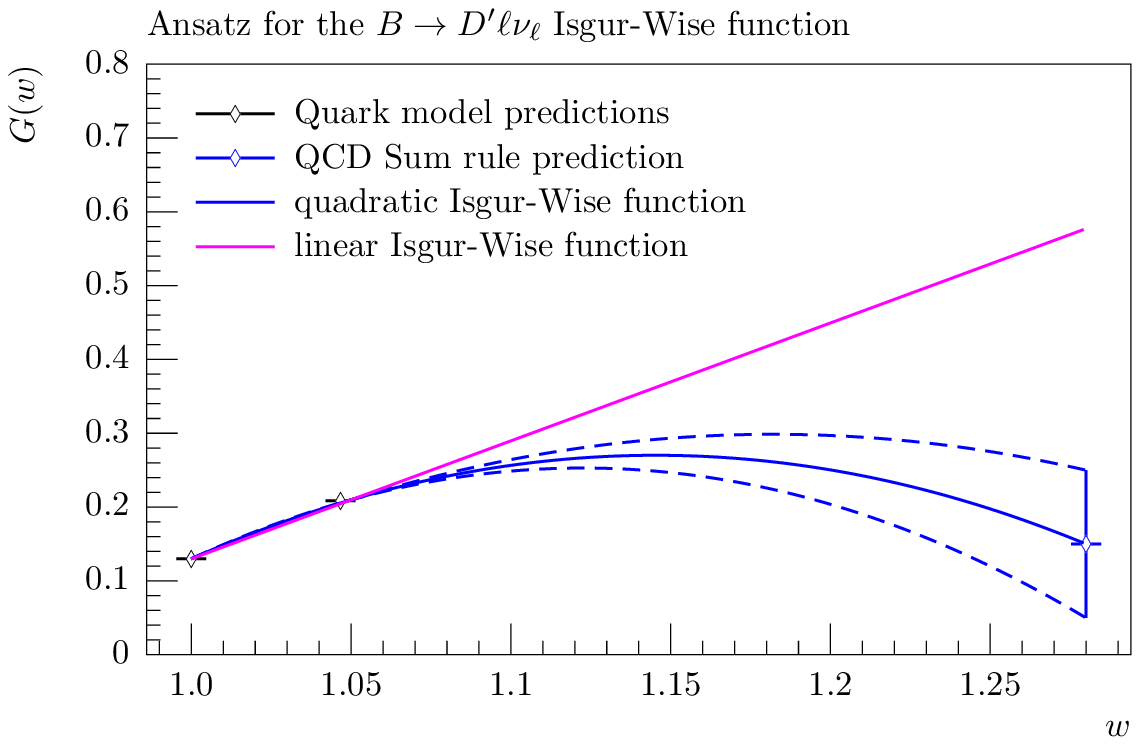}\caption{Isgur-Wise function of the $D^{\prime(*)}$ for $F(w)$ [left] and $G(w)$ [right]}\label{ISGW}
\end{figure}
We obtain from the calculation and the quark model estimate the numbers
\begin{alignat}{2}
         F(1.0) &= 0.10   \qquad  F(1.05) = 0.20 &\qquad  F(w_\text{max}) &= 0.25 \pm 0.15  \nonumber \\
	 G(1.0) &= 0.13 \qquad  G(1.05) = 0.21 &\qquad  G(w_\text{max}) &= 0.15 \pm 0.1\,. \label{ff_calc}
\end{alignat} 
With a linear [quadratic] parameterization for the form factors as
\begin{align}
	G(w)= \beta_0 + (w-1) \beta_1 + \big[(w-1)^2 \beta_2 \big] \\
	 F(w) = \beta_0^{*} + (w-1) \beta_1^{*} + \big[(w-1)^2 \beta_2^{*}\big]
\end{align}
we obtain the parameters the values Eq.~\eqref{ff_calc}, illustrated in Fig.~\ref{ISGW}, 
\begin{alignat}{4}
    \beta_0^* &= 0.10 \,,&\qquad \beta_1^* &= 2.1 \nonumber \\
    \beta_0 &= 0.13 \,,&\qquad \beta_1 &= 1.6 \nonumber\\
    \beta_0^* &= 0.10 \,,&\quad \beta_1^* &= 2.3-2.5 \,,&\quad \beta_2^* &= -(4.2-9.8)  \nonumber \\
    \beta_0 &= 0.13 \,,&\quad \beta_1 &= 1.9-2.0 \,,&\quad \beta_2 &= -(5.1-8.2) \,.
\end{alignat}
Using these interpolations, we obtain the branching fractions in the proposed order of magnitude, which would help to ease or solve the puzzles
\begin{alignat}{2}
  \mathcal{B}\big(B \to (D^{\prime}+D^{\prime *}) \ell \nu_\ell\big) &\sim 1.4 \,\% &\qquad &\text{Linear Interpolation} \nonumber\\
   \mathcal{B}\big(B \to (D^{\prime}+D^{\prime *}) \ell \nu_\ell\big) &\sim (0.3-0.7)\,\%&\qquad &\text{Quadratic Interpolation.} 
\end{alignat}

\section{Discussion}

If future measurements find a substantial $B \to \ensuremath{D^{\prime(*)}} \ell \bar \nu$ decay rate, the precise determination of the branching fraction, the shape of
the $F(w)$ and $G(w)$ functions in Eq.~\eqref{rates}, and data on the
corresponding nonleptonic two-body decays with a pion would be able to test this
picture.  Especially LHCb could contribute by measuring the non-leptonic decay $B\to D^{\prime(*)}\pi$, which is related to the form factors by factorization
	    \begin{equation}
		\Gamma(B\to D^{\prime(*)}\pi) = \frac{3\pi^2 C^2\, |V_{ud}|^2 f_\pi^2}{m_B\, m_{D^{\prime(*)}}}\, \frac{\text{d}\Gamma(B\to D^{\prime(*)}\ell\bar\nu)}{\text{d} w} \Bigg|_{w_\text{max}} 
	    \end{equation}
The strong channel $D'^{(*)} \to D^{(*)} \eta$ could be studied by the present day B-Factories. A considerable radial contribution may also impact other measurements and the theory of semileptonic
decays, e.g., it may yield

\begin{itemize}\vspace*{-4pt}\itemsep 0pt

\item a better understanding of the $b \to c$ background in fully inclusive
$b \to u$ measurements, i.e., lead to a more precise determination of $|
V_{ub}|$;

\item a better understanding of the semileptonic $b \to c$ background in the
exclusive $| V_{cb}|$ measurements using $B \to D^{(*)} \ell\bar\nu$;

\item a better understanding of the missing exclusive contributions to the
inclusive $B\to X_c\ell\bar\nu$ rate, and the lepton energy and hadronic mass
spectrum;

\item a better understanding of the measured $B \to D^{(*)} \tau \bar \nu$ branching fraction 
and its tension with respect to the Standard Model expectation \cite{FrancoSevilla:2011};

\item a more precise determination of the semileptonic branching fractions of
the $s_l^{\pi_l} = \frac12^+$  and $\frac32^+$ states, thus maybe help resolve
the ``1/2 vs.\ 3/2 puzzle";

\item a stronger sum rule bound~\cite{Bigi:1994ga, Kapustin:1996dy,
Dorsten:2003ru,Leibovich:1997em} on the $B\to D^*\ell\bar\nu$
form factor, ${\cal F}(1)$, relevant for the determination of $|V_{cb}|$ from
exclusive decay.

\end{itemize}\vspace*{-4pt}

There are a number of measurements that should be possible using the \mbox{\ensuremath{{\displaystyle B}\!{\scriptstyle A}{\displaystyle B}\!{\scriptstyle AR}}},
Belle, LHCb, and future $e^+e^-$ $B$ factory data samples, which could shed
light on whether this possibility is realized in nature.

\section*{Acknowledgements}
S.T. thanks the organizers for their effort to host this conference and for providing financial support. S.T.~is supported by a DFG Forschungsstipendium under contract no.~TU350/1-1. This work was supported in part by the Director, Office of
Science, Office of High Energy Physics of the U.S.\ Department of Energy under
contract DE-AC02-05CH11231.

\end{document}